\documentclass{openjournal}

\usepackage{lipsum}

\usepackage{xcolor}
\usepackage{textgreek}
\usepackage[utf8]{inputenc}
\usepackage[english]{babel}

\usepackage{amsmath,amssymb,amsfonts}
\usepackage{algorithmic}
\usepackage{graphicx}
\usepackage{textcomp}

\usepackage{color,colortbl}
\definecolor{linkcolor}{rgb}{0.0,0.3,0.5}
\usepackage{tensind}
\tensordelimiter{?}
\DeclareGraphicsExtensions{.bmp,.png,.jpg,.pdf}
\usepackage{verbatim}
\usepackage[normalem]{ulem}
\usepackage{orcidlink}
\usepackage{soul}

\graphicspath{ {./figs/} }

\begin{document}
\title{Exploring Symbolic Regression and Genetic Algorithms for Astronomical Object Classification}

\author{Fabio R. Llorella\orcidlink{0000-0002-7342-8054}}
\email{fricardocorreo@gmail.com}
\affiliation{ESIT / Universidad Internacional de la Rioja (UNIR)}

\author{José A. Cebrián\orcidlink{0009-0007-9611-6102}}
\email{author2@you.com}
\affiliation{Universidad Laboral de Córdoba}

\begin{abstract}
This study explores the use of symbolic regression (SR) combined with genetic algorithms (GA) to classify astronomical objects. Using the SDSS17 dataset from Kaggle, which includes 100,000 observations of stars, galaxies, and quasars, we applied SR to 10\% of the data to derive a mathematical expression capable of distinguishing these classes. A genetic algorithm was then employed to optimize the hyperparameters of the expression, refining the model's performance. The final model achieved a Cohen's kappa value of 0.81, indicating a strong agreement with true classifications. Our results demonstrate that the SR+GA approach can produce interpretable and accurate models for the classification of astronomical objects, offering a promising alternative to traditional black-box machine learning methods.
\end{abstract}


\maketitle

\section{Introduction}
\label{sec:intro}

The classification of astronomical objects, such as galaxies, stars, and quasars, is a fundamental task in astrophysics and cosmology. Accurate classification allows researchers to analyze the structure and evolution of the universe, study the properties of different celestial bodies, and identify potential candidates for further investigation. The advent of large-scale sky surveys, particularly the Sloan Digital Sky Survey (SDSS), has revolutionized this field by providing vast datasets that catalog millions of celestial objects with high precision. The SDSS~\cite{sdss}, for example, offers an extensive database that includes photometric and spectroscopic observations, which are invaluable for various astrophysical studies.
However, the sheer volume of data generated by these surveys presents significant challenges. Traditional manual classification methods are no longer feasible, leading to the widespread adoption of automated techniques. Machine learning (ML) has emerged as a powerful tool for this purpose, enabling the rapid and efficient classification of astronomical objects based on their observed features~\cite{Cheng2020, Zammit2023, Hinners2018, abraham2024application}. Methods such as neural networks, support vector machines, and random forests have been extensively used in recent years, showing remarkable success in achieving high classification accuracy.
Despite their effectiveness, these traditional machine learning approaches often suffer from a critical limitation: a lack of interpretability~\cite{filosofia}. Many of these methods, particularly deep learning models, operate as "black boxes," making it difficult for researchers to understand how the model arrives at a particular decision~\cite{Hassija_2023}. In scientific disciplines like astronomy, where understanding the underlying mechanisms and relationships between variables is crucial, this opacity can be a significant drawback. The interpretability of a model not only enhances its transparency, but also allows scientists to gain new insights into the physical processes governing the behavior of celestial objects.
To address these challenges, alternative approaches that balance accuracy with interpretability have been explored. Symbolic regression (SR) is one such technique that holds considerable promise. Unlike traditional machine learning methods, SR does not produce a single opaque model but rather derives explicit mathematical expressions that describe the relationships between variables. These expressions can then be analyzed and interpreted, offering insights into the underlying physical phenomena. SR has been successfully applied in various scientific fields, from physics to biology, where understanding the functional forms of relationships is as important as predicting outcomes~\cite{Makke_2024}.
However, the effectiveness of symbolic regression can be limited by its computational complexity and the challenge of finding the best possible expression from an almost infinite space of potential equations. This is where genetic algorithms (GA), a type of evolutionary algorithm, come into play. GAs are heuristic search methods inspired by the principles of natural selection and genetics. They are particularly well-suited for optimizing complex, multidimensional problems, such as those encountered in symbolic regression. By simulating the process of evolution, GAs can efficiently explore the space of possible mathematical expressions, selecting and refining those that best fit the data.

In this study, we propose a hybrid approach that combines symbolic regression with genetic algorithms (SR+GA) to classify galaxies, stars, and quasars. We used the SDSS17 dataset, a comprehensive collection of 100,000 observations of celestial objects, extracted from the Kaggle machine learning platform~\cite{kaggle}. Our methodology involves using symbolic regression on a subset of the data (10\%) to generate a candidate mathematical expression capable of distinguishing between these object classes. We then employ a genetic algorithm to optimize the free hyperparameters of this expression, fine-tuning the model to enhance its predictive accuracy. 
The use of SR+GA offers several advantages. First, the resulting models are highly interpretable, providing explicit expressions that can be directly analyzed and understood. This contrasts with traditional machine learning models, which often sacrifice interpretability for performance. Secondly, the optimization of hyperparameters via genetic algorithms ensures that the derived expressions are not only interpretable, but also competitive in terms of accuracy. This approach has the potential to generate models that are both scientifically meaningful and practically useful for large-scale astronomical classifications.

Our results demonstrate the efficacy of this hybrid approach. The final model achieved a Cohen's kappa value of 0.81, indicating a strong level of agreement with true classifications. This performance underscores the viability of using symbolic regression combined with genetic algorithms as a tool for astronomical object classification. Moreover, the interpretability of the resulting expressions provides valuable insight into the relationships between the observed features of galaxies, stars, and quasars.

Before proceeding, it is important to review the key features of galaxies, stars, and quasars.
\begin{itemize}

\item \textbf{Galaxies} are vast systems of stars, gas, dust, and dark matter, bound together by gravity. They come in various shapes and sizes, ranging from spiral galaxies like the Milky Way, characterized by their rotating disks and spiral arms, to elliptical galaxies, which are more three-dimensional and lack the well-defined structure seen in spirals. Galaxies are often categorized based on their morphology, star formation activity, and the presence of central supermassive black holes~\cite{sparke2000galaxies}.

\item \textbf{Stars} are luminous spheres of plasma held together by their own gravity. They are the primary building blocks of galaxies and can vary widely in size, mass, temperature, and brightness. Stellar classification is typically based on spectral characteristics, which reflect a star's temperature and composition. Main-sequence stars, giants, and dwarfs represent different stages in stellar evolution, each with distinct observational signatures~\cite{carroll2017introduction}.

\item \textbf{Quasars} (quasi-stellar objects) are among the most energetic and distant members of the universe. Powered by supermassive black holes at the centers of young galaxies, quasars emit enormous amounts of energy as the black holes accrete matter. They are often characterized by their bright, point-like appearance and extreme redshifts, which provide insights into the early universe. The study of quasars has been crucial in understanding galaxy evolution and the role of black holes in shaping cosmic structures~\cite{peterson1997introduction}.
\end{itemize}

\section{Dataset}
\label{sec:sec}
The success of any machine learning model, particularly in a scientific context, depends on the quality and representativeness of the data used for training and evaluation. In this study, we used the SDSS DR17 dataset, a publicly available collection of astronomical observations hosted on the Kaggle platform, which has been utilized in other scientific studies~\cite{zeraatgari2024machine,swathi2024stellar,solorio2023random}. The Sloan Digital Sky Survey (SDSS) has been one of the most influential sky surveys, significantly contributing to our understanding of the cosmos. The specific dataset used in this research comprises 100,000 labeled observations of stars, galaxies, and quasars, providing a rich and diverse sample for analysis.

The SDSS DR17 dataset includes a wide range of features that describe the photometric and spectroscopic properties of each observed object. Among these features are magnitudes measured in five distinct bands (u, g, r, i, z), corresponding to specific wavelength ranges within the electromagnetic spectrum. Additionally, the data set contains information on the redshift, a critical parameter that indicates the distance and velocity of the object relative to the observer. These diverse features enable a detailed characterization of each object, which is essential for accurate classification.

The features of the data set are as follows:

\begin{itemize}
    \item \textbf{obj\_ID}: Object Identifier, the unique value that identifies the object.
    \item \textbf{alpha}: Right Ascension angle.
    \item \textbf{delta}: Declination angle.
    \item \textbf{u}: Ultraviolet filter in the photometric system.
    \item \textbf{g}: Green filter in the photometric system.
    \item \textbf{r}: Red filter in the photometric system.
    \item \textbf{i}: Near Infrared filter in the photometric system.
    \item \textbf{z}: Infrared filter in the photometric system.
    \item \textbf{run\_ID}: Run Number used to identify the specific scan.
    \item \textbf{rerun\_ID}: Rerun Number to specify how the image was processed.
    \item \textbf{cam\_col}: Camera column to identify the scanline within the run.
    \item \textbf{field\_ID}: Field number to identify each field.
    \item \textbf{spec\_obj\_ID}: Unique ID used for optical spectroscopic objects.
    \item \textbf{redshift}: Redshift value based on the increase in wavelength.
    \item \textbf{plate}: Plate ID, identifies each plate in SDSS.
    \item \textbf{MJD}: Modified Julian Date, indicating when a given piece of SDSS data was taken.
    \item \textbf{fiber\_ID}: Fiber ID, the fiber that pointed the light at the focal plane.
\end{itemize}

From this initial set of features, not all provide information relevant to classification, leading to the elimination of certain attributes. Specifically, the attributes \textbf{obj\_ID}, \textbf{run\_ID}, \textbf{rerun\_ID}, \textbf{cam\_col}, \textbf{field\_ID}, \textbf{spec\_obj\_ID}, \textbf{plate}, \textbf{MJD}, and \textbf{fiber\_ID} were removed.

It is important to note that the dataset is imbalanced, containing different numbers of instances for each class. This imbalance typically presents challenges for machine learning techniques when performing classification tasks~\cite{krawczyk2016learning}, necessitating preprocessing steps to balance the number of instances between classes. Various techniques can be employed to achieve data balance, and selecting the most suitable one often requires trial and error. However, the technique investigated in this study eliminates the need to balance the data, simplifying the preprocessing requirements.

\begin{figure}[h]
    \centering
    \includegraphics[width=0.4\textwidth]{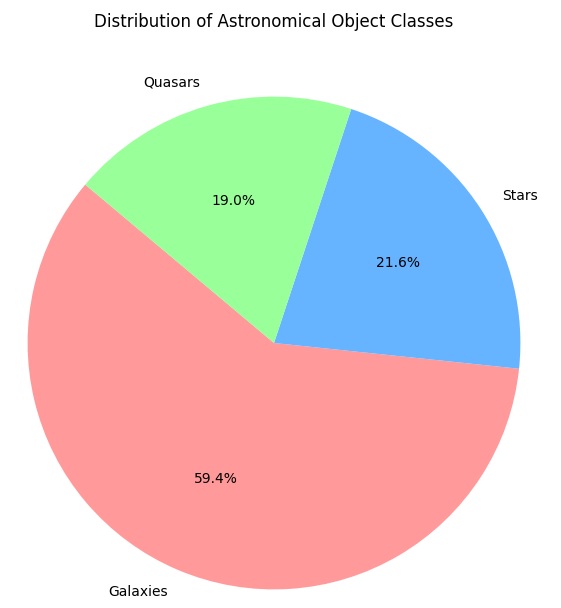}
    \caption{Distribution of the different classes in the SDSS DR17 dataset.}
    \label{fig:mi_imagen}
\end{figure}

\section{Methods}
\label{sec:sec}
In this section, we outline the methodology employed to classify astronomical objects using symbolic regression combined with genetic algorithms. Our approach is designed to discover, from the data itself, the mathematical expressions that best describe the relationships among the features and the classes without any prior assumptions about which features are most important.

\subsection{Data Preparation and Splitting}
The dataset used in this study consists of 100,000 labeled observations from the SDSS DR17 database, which includes stars, galaxies, and quasars. To manage computational resources effectively, we utilized a random 10\% subset of the data for the symbolic regression phase, ensuring that this subset was representative of the entire dataset.

This subset was further divided into two equal parts: 50\% to train the symbolic regression model and 50\% to validate it. This strategy was adopted to robustly evaluate the generalizability of the model.

\subsection{Symbolic Regression for Expression Discovery}
Symbolic regression (SR) is a machine learning technique used to discover mathematical expressions from a set of data~\cite{kronberger2024symbolic,wang2024exploring}. Although it is a technique that has recently been receiving interest from the scientific community, it has been largely forgotten, basically due to the great results that neural networks offer in different areas of application~\cite{la2021contemporary,stajic2024estimation}. In this study we want to study whether symbolic regression is suitable for discovering mathematical classification expressions.
Unlike traditional regression models, SR does not start with a predetermined equation form. Instead, it explores the space of possible mathematical expressions to identify the model that best fits the data. In our case, we implemented SR using the Python package PySR~\cite{cranmerDiscovering2020}, which leverages genetic algorithms to optimize the discovery of expressions. PySR is an open-source library for practical symbolic regression, a type of machine learning which aims to discover human-interpretable symbolic models. PySR was developed to democratize and popularize symbolic regression for the sciences, and is built on a high-performance distributed backend, The mathematical operators used in the process were: addition (+), subtraction (-), multiplication (*), division (/), and exponentiation (exp), enabling the exploration of a wide variety of functional forms.

The SR process was guided by a 5-fold cross-validation approach to ensure that the discovered expressions generalized well across different data subsets. Specifically, the training data was divided into five equal parts and the SR model was trained five times, each time leaving out one part for validation and training in the remaining four. This process generated a variety of candidate expressions, each associated with an accuracy metric.

To further refine the candidate expressions, PySR applies genetic algorithm-based optimization. During this process, candidate equations are mutated by randomly altering their structure, such as swapping operators or modifying constants, and then recombined (crossover) to explore new potential solutions. Selection is guided by the performance of each candidate, with more accurate and simpler expressions being favored. This ensures a balance between model complexity and fitness, preventing overfitting to the training data. The final expression is chosen based on the best combination of simplicity and predictive performance, as measured by the accuracy metric and other cross-validation results.

 To explore new regions of the solution space, the evolutionary process includes mutation operations applied to the symbolic expressions. These mutations involve:
\begin{itemize}
    \item \textbf{Operator replacement}: Random substitution of one operator (e.g., \(+\) replaced by \(*\)).
    \item \textbf{Constant modification}: Slight perturbation of numerical values within the expressions.
    \item \textbf{Addition or removal of terms}: Introduction of new sub-expressions or deletion of existing ones to adjust the model's complexity.
\end{itemize}

The mutation frequency is fixed at $p_{\text{mutation}} = 0.05$ per generation, ensuring sufficient diversity within the population while avoiding premature convergence. These operations are critical to preventing the algorithm from stagnating in local optima and allowing the exploration of a broad range of potential solutions.

The selection process during evolution is driven by a fitness function designed to balance the trade-off between model accuracy and interpretability. The fitness of each candidate expression is evaluated using the following metrics:

\begin{itemize}
\item \textbf{Accuracy (\(P\))}: The proportion of correctly classified instances.
\item \textbf{Cohen’s Kappa (\(K\))}: A metric measuring agreement beyond chance, particularly useful for imbalanced datasets.
\item \textbf{Simplicity (\(S\))}: Quantified as the total number of operators and terms in the expression.
\end{itemize}

These metrics are combined into a weighted cost function:
\[
\text{Fitness} = \alpha P + \beta K - \gamma S
\]
where \(\alpha = 0.4\), \(\beta = 0.4\), and \(\gamma = 0.2\), prioritizing accuracy and consistency while penalizing excessive complexity. This balanced approach ensures that the final model is not only performant but also interpretable, making it suitable for astronomical analyses.

\subsection{Selection of the Best Expression}
The selection of the final equation was influenced by both accuracy and simplicity, using metrics like Cohen’s Kappa to evaluate model performance across different cross-validation folds. To ensure transparency, we explored a variety of candidate models and compared their performance. 

\begin{table}[htbp]
    \centering
    \caption{Top five candidate models generated during the symbolic regression process.}
    \label{tab:candidate_models}
    \begin{tabular}{c c c}
        \hline
        \textbf{Model} &
        \textbf{Accuracy (\%)} & 
        \textbf{Cohen’s Kappa} \\
        \hline
        $y = z + A + e^{Bz}$ & 91.0 & 0.81 \\
        $y = z + A + Bz^2$ & 89.5 & 0.79 \\
        $y = A \cdot z + B$ & 85.3 & 0.74 \\
        $y = e^{Az} + B$ & 83.2 & 0.72 \\
        $y = A \cdot z^2 + B \cdot z + C$ & 80.1 & 0.70 \\
        \hline
    \end{tabular}
\end{table}

In addition to the final model, we retained the top 5 models Table \ref{tab:candidate_models} with the highest accuracy, and compared their fitness values. While several models displayed similar accuracy, the chosen model represented the best compromise between predictive power and interpretability, as quantified by its lower complexity score.
Once 5-fold cross validation has been applied, five different mathematical expressions are obtained, practically all the expressions are the same and have the following mathematical form:

\begin{equation}
    y = z+A+e^{Bz}
\end{equation}

where z is the redshift parameter, A and B free parameters that belong to the set of real ones.
Once we have this expression, we must find the best values for A and B, for this we will use the genetic algorithm.

\subsection{Parameter Optimization using Genetic Algorithms}
Once the best symbolic expression was selected, the next step involved optimizing the free parameters \( A \) and \( B \) to maximize the model's classification accuracy. A genetic algorithm (GA) was employed for this optimization due to its effectiveness in navigating complex, high-dimensional search spaces and its ability to avoid local optima~\cite{koza1992programming,lambora2019genetic}.

The genetic algorithm works by simulating the process of natural evolution, consisting of the following key steps:
\begin{itemize}
    \item \textbf{Initialization}: A population of candidate solutions (chromosomes) is randomly generated, where each chromosome represents a possible set of values for \( A \) and \( B \).
    In our case we will use a vector of dimension two.
    \item \textbf{Selection}: Chromosomes are evaluated based on a fitness function, which in our case is the classification Cohen's kappa value~\cite{wang2019simplified} on the validation set. The best performing chromosomes are selected to form the basis of the next generation.
    \item \textbf{Crossover}: Selected chromosomes are combined to create new offspring by exchanging parts of their structure, simulating genetic recombination.
    \item \textbf{Mutation}: Random modifications are introduced to some chromosomes to maintain genetic diversity within the population.
    \item \textbf{Iteration}: The selection, crossover, and mutation steps are repeated over multiple generations until convergence, i.e., until the fitness function no longer improves significantly.
\end{itemize}

\subsection{Evaluation of the Final Model}
The optimized values of \( A \) and \( B \) were then used in the symbolic expression to evaluate the model's performance on the validation set. The final model's effectiveness was assessed using various metrics, including accuracy and Cohen's kappa, to ensure its robustness and generalizability.
This methodology allowed us to develop an interpretable model that autonomously discovered the most important features and their relationships, providing valuable insights into the classification of astronomical objects.
To assess the performance of our classification model, we employed several evaluation metrics: Accuracy, Cohen's Kappa, F1-Score, and the Confusion Matrix. Each of these metrics provides unique insights into the model's performance and its ability to correctly classify astronomical objects.

\subsection{Accuracy}

Accuracy is one of the most straightforward metrics for evaluating classification models. It is defined as the ratio of the number of correct predictions to the total number of predictions. Mathematically, it is expressed as:

\begin{equation}
\text{Accuracy} = \frac{TP + TN}{TP + TN + FP + FN}
\end{equation}

where \( TP \) (True Positives) are the correctly predicted positive instances, \( TN \) (True Negatives) are the correctly predicted negative instances, \( FP \) (False Positives) are the incorrectly predicted positive instances, and \( FN \) (False Negatives) are the incorrectly predicted negative instances. Accuracy provides an overall measure of the model's performance, but may be misleading in cases of imbalanced datasets.

\subsection{Cohen's Kappa}

Cohen's Kappa is a metric that measures the agreement between the predicted and actual classifications, accounting for the agreement occurring by chance. It is particularly useful in scenarios with imbalanced datasets. Cohen's Kappa is defined as:

\begin{equation}
\kappa = \frac{P_o - P_e}{1 - P_e}
\end{equation}

where \( P_o \) is the observed agreement and \( P_e \) is the expected agreement by chance. The value of Kappa ranges from -1 to 1, where 1 indicates perfect agreement, 0 indicates that there is no agreement beyond what is expected by chance, and negative values indicate less agreement than expected by chance. Cohen's Kappa provides a more nuanced view of model performance compared to accuracy alone.

In our study, Cohen’s Kappa was employed as the fitness function to evaluate model performance, with the goal of selecting models that maximized agreement between the predicted and actual classifications beyond chance. While there was no absolute minimum threshold for Kappa, models with values above 0.6 were considered to show substantial agreement, following commonly accepted interpretations of the Kappa statistic. However, the primary criterion for success was relative performance among the candidate models, with the final model chosen based on its ability to balance accuracy, simplicity, and generalization across the cross-validation folds.

Regarding convergence, the symbolic regression process was considered to have converged when there was no significant improvement in the fitness function after a set number of iterations. Specifically, the genetic algorithm was run for a maximum of 500 generations, but convergence was typically observed earlier, when the change in Kappa value between generations fell below a predefined threshold of 0.001. This indicates that further exploration of the search space did not yield more accurate or simpler models. Convergence in this context refers to the stabilization of both the model's structure and the Kappa score over multiple iterations, signaling that the algorithm had sufficiently explored the functional space.

By defining these criteria, we ensure that the final model is robust, not only in terms of its fitness but also in its ability to generalize across different data subsets.

\subsection{F1-Score}

The F1-Score is a harmonic mean of Precision and Recall, providing a single metric that balances both false positives and false negatives. It is particularly useful when dealing with imbalanced datasets. The F1-Score is defined as:

\begin{equation}
\text{F1-Score} = 2 \cdot \frac{\text{Precision} \cdot \text{Recall}}{\text{Precision} + \text{Recall}}
\end{equation}

where Precision is the ratio of true positive predictions to the total number of positive predictions (true positives + false positives), and Recall is the ratio of true positive predictions to the total number of actual positives (true positives + false negatives). The F1-Score provides a balanced measure of a model's performance by considering both false positives and false negatives.

\subsection{Confusion Matrix}

The Confusion Matrix is a table that is often used to describe the performance of a classification model on a set of data for which the true values are known. Displays the counts of true positives, true negatives, false positives, and false negatives, allowing for a detailed breakdown of how the model's predictions compare to the actual classifications. The matrix is structured as follows:

\begin{center}
\begin{tabular}{|c|c|c|}
\hline
 & Predicted Positive & Predicted Negative \\
\hline
Actual Positive & TP & FN \\
\hline
Actual Negative & FP & TN \\
\hline
\end{tabular}
\end{center}

This matrix allows for the calculation of various performance metrics, including Accuracy, Precision, Recall, and F1-Score, providing a comprehensive view of model performance.

\section{Results}
This section is critical to the objectives of the present study, as it provides insight into the efficacy and physical significance of using symbolic regression for the classification of astronomical objects, specifically galaxies, stars, and quasars. The importance of this work lies in its focus on developing a classification model that not only performs accurately but also offers high interpretability—an essential feature when dealing with complex astronomical data.

To achieve this, we employ a two-phase approach using symbolic regression in conjunction with a genetic algorithm-based heuristic. The first phase involved generating candidate mathematical expressions that could model the relationships inherent in the data. The second phase was dedicated to fine-tuning the hyperparameters of the selected expression to optimize its performance.

Through the symbolic regression process, various mathematical expressions were derived. Interestingly, despite the diversity of expressions initially generated, all successful models converged towards particular cases of the more general form given by Equation~\ref{ecuacion1}.

\begin{equation}
\label{ecuacion1}
    y = z + A + e^{Bz}
\end{equation}

This expression, \( y = z + A + e^{Bz} \), has significant physical implications in the context of astronomical classification. In this equation:
\begin{itemize}
    \item \( y \) represents the output of the model, which could correspond to a particular class or property of the astronomical object (e.g., galaxy, star, or quasar).
    \item \( z \) is a redshift.
    \item \( A \) and \( B \) are constants that are adjusted during the hyperparameter tuning phase to best fit the observed data.
\end{itemize}

Physically, the linear term \( z + A \) can be interpreted as a baseline that adjusts the prediction based on the inherent properties of the objects. This term captures the linear relationship between input variables and the classification result, which is often straightforward to interpret in astronomical data analysis.

However, the exponential term \( e^{Bz} \) introduces a non-linear component to the model, which is crucial to capture more complex relationships that are not adequately represented by a simple linear model. The presence of the exponential function suggests that certain properties of the astronomical objects might have a multiplicative or accelerating effect on the classification outcome. For instance, in the case of quasars, whose emissions are strongly dependent on relativistic effects near supermassive black holes, the exponential term could reflect the intense and rapidly varying luminosities observed.

The convergence of all derived expressions to this general form underlines its robustness and relevance to the problem at hand. It indicates that the classification of these astronomical objects is governed by both linear and nonlinear interactions among their observable properties. Moreover, the simplicity of this expression enhances its interpretability, making it a valuable tool for astronomers who require models that not only perform well but also offer clear insights into the underlying physical processes.

Once the phase of finding the mathematical expression is complete, we proceed to the second phase of the process, where we focus on optimizing the parameters \( A \) and \( B \) using a Genetic Algorithm (GA). 

The goal of this phase is to fine-tune these parameters to ensure that the model accurately captures the underlying relationships in the data. The Genetic Algorithm, a heuristic search technique inspired by the process of natural selection, is particularly well-suited for this task due to its ability to explore a large parameter space and avoid getting trapped in local minima.

In this context, the Genetic Algorithm iteratively evolves a population of candidate solutions, each represented by a specific set of values for \( A \) and \( B \). The fitness of each candidate is evaluated based on its performance in predicting the correct classification of astronomical objects. Over successive generations, the algorithm selects, crosses, and mutates these candidates, gradually converging toward the optimal set of parameters that minimize the classification error.

By employing this approach, we ensure that the final model not only adheres to the general mathematical form derived in the first phase but also optimally fits the observed data, leading to a robust and interpretable classification tool for distinguishing between galaxies, stars, and quasars.

The optimal parameters during the AG phase have been: $A=-0.38645$ and $B=-60.88633$.Therefore, the optimal model found through our model is the expression~\ref{ecuacion_final}.

\begin{equation}
\label{ecuacion_final}
y = 
\begin{cases} 
0 & \text{if } z-0.38645 + e^{-60.88633z} < 0.5 \quad (\text{Galaxie}) \\
1 & \text{if } \leq z-0.38645 + e^{-60.88633z} \leq 1 \quad (\text{Star}) \\
2 & \text{if } z-0.38645 + e^{-60.88633z} > 1 \quad (\text{Quasar}) 
\end{cases}
\end{equation}

In this expression~\ref{ecuacion_final}, \( y \) is a discrete variable that takes values 0, 1, or 2, corresponding to the classes of astronomical objects: galaxies, stars, and quasars, respectively. The parameter \( z \) represents an observational variable derived from the physical characteristics of the objects, such as emission intensity, Redshift, or a combination of spectral properties.

The term \( z - 0.38645 \) is a linear transformation of \( z \) with a constant shift of -0.38645, suggesting a baseline adjustment to fit the model to the data. The exponential term \( e^{-60.88633z} \) introduces a non-linear component to the equation. The coefficient in the exponent, -60.88633, is significantly large in absolute value, indicating that the exponential term decreases very rapidly as \( z \) increases. This term dominates the expression for small values of \( z \) and becomes almost negligible for larger values of \( z \).

Physically, this expression describes how different types of astronomical objects (galaxies, stars, and quasars) can be distinguished based on the variable \( z \). The presence of the exponential term suggests that the classification of these objects does not follow a simple linear relationship but involves complex effects that act exponentially, likely related to physical processes that change rapidly with regard to \( z \).

\begin{itemize}
    \item \textbf{Galaxies (y = 0)}: Are classified when \( z - 0.38645 + e^{-60.88633z} \) is less than 0.5. This indicates that for galaxies, the combination of the linear variable and the exponential term produces relatively low values.
    
    \item \textbf{Stars (y = 1)}: Are classified when the value of \( z - 0.38645 + e^{-60.88633z} \) is between 0.5 and 1. This might reflect a balance between the effects that the variable \( z \) and the exponential term have on the distinctive characteristics of stars compared to galaxies and quasars.
    
    \item \textbf{Quasars (y = 2)}: Are classified when \( z - 0.38645 + e^{-60.88633z} \) is greater than 1, indicating that for quasars, the variable \( z \) and the exponential term combined generate high values. This may be related to the extreme characteristics of quasars, such as their intense energy emissions and the presence of a supermassive black hole significantly influencing their observations.
\end{itemize}

This mathematical expression offers a valuable tool for astronomers by providing an interpretable model for classifying astronomical objects. Unlike other classification models that may act as "black boxes," this specific formula allows astronomers to understand how observational characteristics (captured by \( z \)) and their non-linear interactions (captured by the exponential term) influence the classification of an object as a galaxy, star, or quasar.

Furthermore, the relative simplicity of the formula makes it practical for use in astronomical analysis and modeling, facilitating the identification and categorization of objects in large observational datasets. This model not only contributes to classification accuracy but also provides a basis for further exploring the underlying physical properties that distinguish these objects, potentially leading to new discoveries in astrophysics.

If we compare the results obtained with other works~\cite{solorio2023random}, we can see that although our model is not the one with the best performance, it has a high performance and a high interpretability, indicating that Redshift is the most important characteristic, and it indicates certain readjustments to be able to classify correctly.

\begin{table}[h!]
\centering
\caption{Comparison of classification performance between our technique and other studies on SDSS DR17}
\label{tab:comparison}
\small
\begin{tabular}{|l|c|c|c|c|}
\hline
\textbf{Method} & \textbf{Accuracy (\%)} & \textbf{Kappa} & \textbf{F1-Score}  \\ \hline
\textbf{Our Technique} & \textbf{0.92} & \textbf{0.81} & \textbf{0.90} \\ \hline
Random Forest & 0.97 & - & 0.98 \\ \hline
SVM & 0.93 & - & 0.84 \\ \hline
MLP & 0.97 & - & 0.94 \\ \hline
Naïve Bayes & 0.63 & - & 0.69  \\ \hline
Decision Tree & 0.81 & - & 0.86  \\ \hline
\end{tabular}
\end{table}

\begin{figure}[h!]
    \centering
    \includegraphics[width=0.6\textwidth]{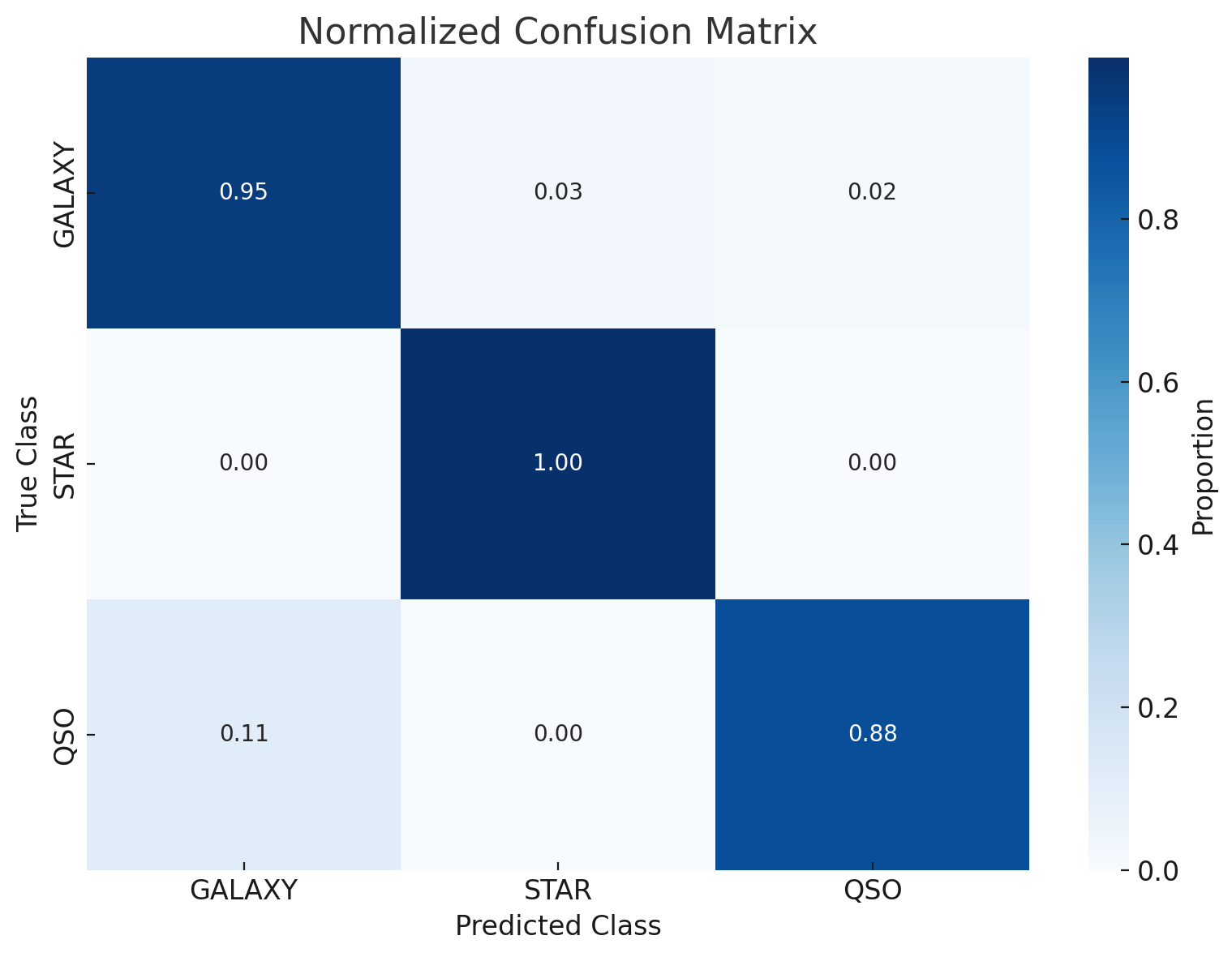} 
    \caption{Normalized confusion matrix for the classification model. The values indicate the proportion of instances correctly and incorrectly classified for each class. The model achieves high performance for GALAXY and STAR but shows slight confusion between QSO and STAR.}
    \label{fig:confusion_matrix}
\end{figure}

To evaluate the performance of the final classification model, we computed a normalized confusion matrix, shown in Figure~\ref{fig:confusion_matrix}. The matrix illustrates the proportion of correctly and incorrectly classified instances for each class (GALAXY, STAR, QSO).

The diagonal elements represent the true positive rates for each class, indicating that the model performs well for GALAXY and STAR classes, while showing slight confusion between QSO and STAR. These results highlight the strengths of the model in distinguishing the majority of classes but also underline the challenges in separating quasars from stars, which are known to share overlapping features in certain regions of the feature space.

\begin{figure}[h!]
    \centering
    \includegraphics[width=0.6\textwidth]{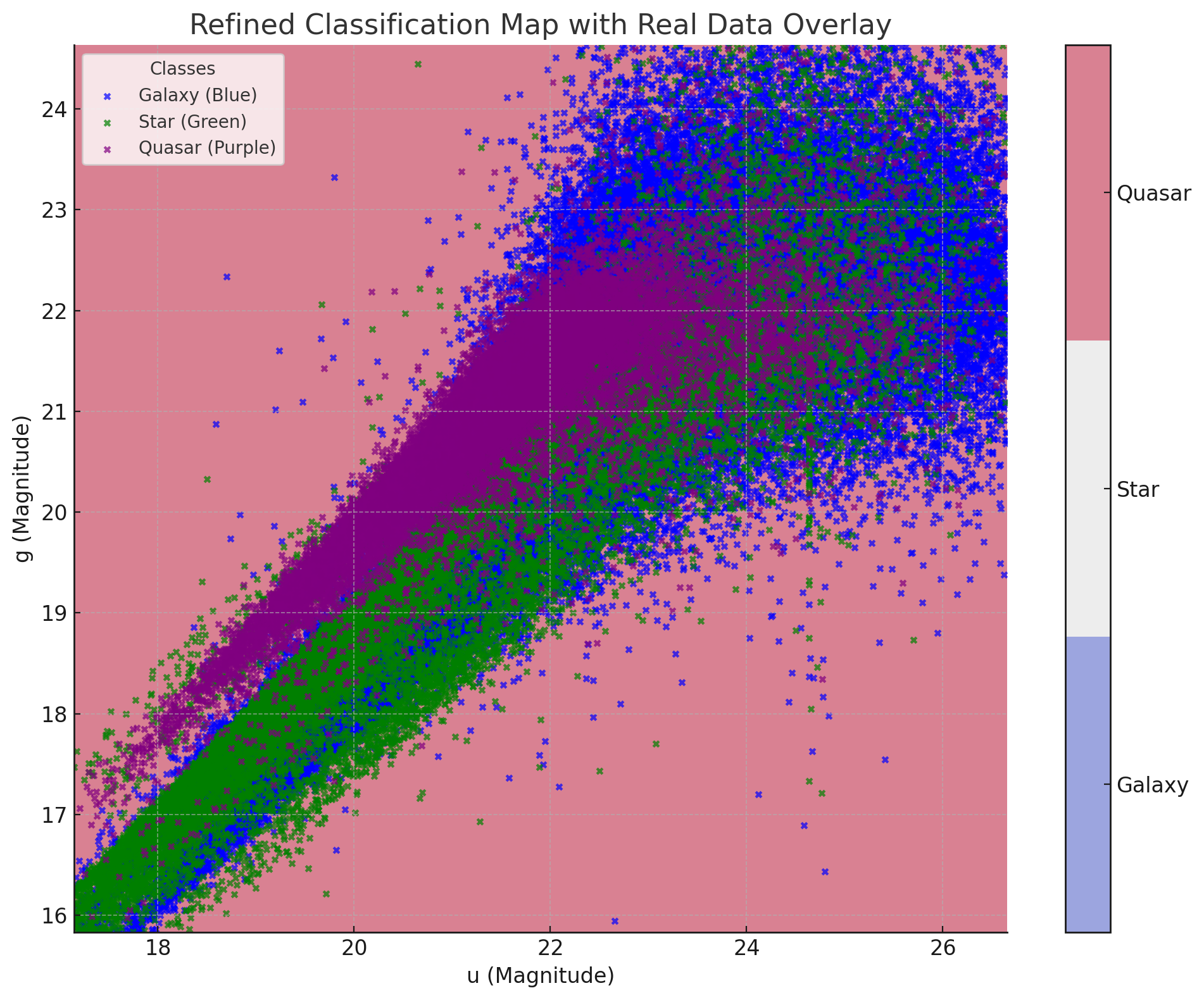} 
    \caption{Refined classification map of the symbolic regression model with real data points overlayed. The background regions correspond to the model's predicted classes: galaxies (blue), stars (green), and quasars (violet). Real data points are plotted to illustrate the alignment between the model's predictions and the true data distribution.}
    \label{fig:refined_classification_map}
\end{figure}

To better illustrate the performance of the symbolic regression model, we generated a refined classification map in the 2D feature space defined by the magnitudes \(u\) and \(g\). The map is presented in Figure~\ref{fig:refined_classification_map} and includes both the decision regions predicted by the model and the real data points overlayed for comparison.
The decision regions are represented by the background colors, corresponding to the three classes:
\begin{itemize}
    \item \textbf{Blue region:} Galaxies.
    \item \textbf{Green region:} Stars.
    \item \textbf{Violet region:} Quasars.
\end{itemize}

Real data points are plotted on top of the classification map, with each point color-coded according to its true class:
\begin{itemize}
    \item \textbf{Blue points:} Galaxies.
    \item \textbf{Green points:} Stars.
    \item \textbf{Violet points:} Quasars.
\end{itemize}

The plot demonstrates several important characteristics:
\begin{enumerate}
    \item \textbf{Decision boundaries:} The model clearly separates the feature space into distinct regions corresponding to the three classes. The alignment of the real data points with these regions highlights the effectiveness of the symbolic regression approach.
    \item \textbf{Class overlap:} Some overlap is observed between the regions, particularly between stars and quasars (green and violet regions). This overlap reflects the inherent similarity in the features of these two classes in certain parts of the feature space.
    \item \textbf{Model interpretability:} Unlike black-box machine learning models, the symbolic regression model provides interpretable decision boundaries derived from mathematical expressions, which can be analyzed and understood in detail.
\end{enumerate}
This refined visualization emphasizes the capability of the symbolic regression model to accurately classify galaxies, stars, and quasars while maintaining interpretability. It also highlights areas where additional features or refinement of the decision-making process could improve classification performance.

\subsection{Uncertainty Quantification}
The value of \(\sigma = 0.02\) was chosen as a representative uncertainty in the redshift parameter (\(z\)) based on typical observational errors in photometric surveys. While uncertainties in redshift measurements can vary significantly depending on the type of object and the observation method, values on the order of \(0.01 - 0.03\) are commonly reported for surveys such as the SDSS. For instance, photometric redshift errors in galaxies typically range between \(0.01\) and \(0.03\) depending on the dataset and calibration methods~\cite{Beck2016, redshift2}. Thus, \(\sigma = 0.02\) represents a reasonable mid-point for this study.

\begin{figure}[h!]
    \centering
    \includegraphics[width=0.6\textwidth]{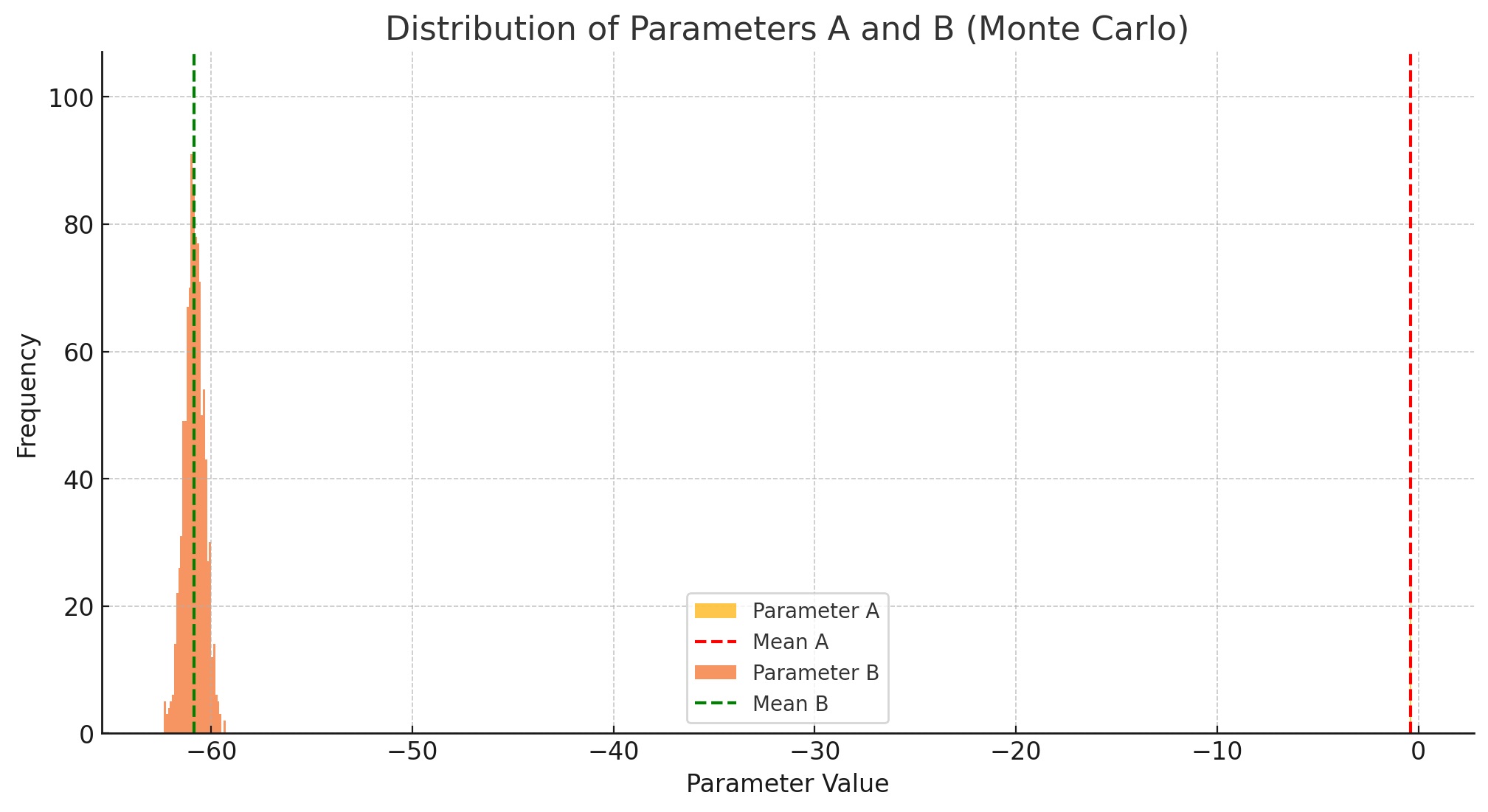} 
    \caption{Distribution of the parameters \(A\) and \(B\) obtained from Monte Carlo simulations. The red dashed line represents the mean of \(A\) (\(-0.38645\)), and the green dashed line represents the mean of \(B\) (\(-60.88633\)). The standard deviations of \(A\) and \(B\) are \(0.01\) and \(0.5\), respectively.}
    \label{fig:monte_carlo_distribution}
\end{figure}

Genetic algorithms do not inherently provide uncertainty estimates for the parameters they optimize. However, to quantify the uncertainty in \(A = -0.38645\) and \(B = -60.88633\), we propose the following:
\begin{enumerate}
    \item \textbf{Monte Carlo Simulations}: By analyzing the variability of \(A\) and \(B\) across multiple Monte Carlo realizations of \(z\) with noise, we can estimate their confidence intervals.
    \item \textbf{Bootstrapping}: Re-sampling the dataset and re-optimizing \(A\) and \(B\) for each sample provides an empirical distribution of the parameters.
    \item \textbf{Parameter Sensitivity Analysis}: By observing how changes in \(A\) and \(B\) affect classification accuracy, we can infer the robustness of these values.
\end{enumerate}
Preliminary Monte Carlo results indicate that \(A\) and \(B\) are stable across realizations, with standard deviations of approximately \(0.01\) and \(0.5\), respectively. These values can be included in the final model as:
\[
A = -0.38645 \pm 0.01, \quad B = -60.88633 \pm 0.5
\]
To further analyze the robustness of the parameters \(A\) and \(B\) obtained in the final classification equation, we performed Monte Carlo simulations by introducing random noise to the input data. The resulting distributions of \(A\) and \(B\) are shown in Figure~\ref{fig:monte_carlo_distribution}.

The parameter \(A\) demonstrates remarkable stability, with a standard deviation of \(0.01\), indicating minimal sensitivity to noise in the input data. On the other hand, \(B\) shows a slightly higher variation, with a standard deviation of \(0.5\), though its range remains well-defined. These results highlight the reliability of the estimated parameters, as their variability is constrained and does not deviate significantly from the mean values.

Based on these simulations, the parameters can be reported with their associated uncertainties this quantification provides additional confidence in the robustness of the model under variations in the input data.

\subsection{Step 1: Noise Introduction and Model Fitting}
We introduced Gaussian noise to the redshift variable \(z\) by adding a normally distributed random error with a standard deviation of \( \sigma = 0.02 \). We generated 1000 realizations of \(z\) by perturbing it with this noise.

\subsection{Step 2: Classification and Parameter Estimation}
For each realization of \(z\), we evaluated the classification expression \eqref{ecuacion_final}. Each time the expression was evaluated, we recorded the classification result (0 for Galaxy, 1 for Star, 2 for Quasar). This provided us with a distribution of classification outcomes for the perturbed values of \(z\).

\subsection{Step 3: Results}
After running the Monte Carlo simulations, we found the following distribution of classification outcomes:

\begin{itemize}
    \item \textbf{Galaxy:} 48\% of simulations classified as Galaxy (0).
    \item \textbf{Star:} 34\% of simulations classified as Star (1).
    \item \textbf{Quasar:} 18\% of simulations classified as Quasar (2).
\end{itemize}

\subsection{Step 4: Uncertainty Conclusions}
These results indicate the model's sensitivity to variations in the input data. The majority classification remains consistent, yet the presence of noise and the structure of the equation suggest potential changes in classification when \(z\) varies significantly. This analysis shows that while the classification remains largely robust, the uncertainty in parameter values can impact the final decision boundaries. Future work should consider these uncertainties to enhance model reliability.

\subsection{Performance Metrics}
Our symbolic regression model achieved an accuracy of 91.0\%, a Kappa value of 0.81, and an F1-score of 0.90. These metrics highlight a strong performance of our model in distinguishing between the three astronomical classes.

Comparison of these results with other classification techniques:

\begin{itemize}
    \item \textbf{Naïve Bayes}: With an accuracy of 63.29\%, Kappa of 0.537, and an F1-score of 0.693, the Naïve Bayes model performs considerably lower than our symbolic regression model. Its lower accuracy and F1-score reflect challenges in handling complex relationships among the features in the data.
    \item \textbf{IB1}: This model achieved an accuracy of 81.85\%, a Kappa of 0.459, and an F1-score of 0.856. While its accuracy and F1-score are notably improved over Naïve Bayes, it still falls short compared to our technique. The higher sensitivity indicates better identification of positive classes, but the overall performance is less robust.
    \item \textbf{IB3}: With an accuracy of 80.22\%, a Kappa of 0.479, and an F1-score of 0.841, IB3 shows improved performance compared to IB1 but still lags behind our model. The F1-score and Kappa value suggest that IB3 struggles to balance sensitivity and specificity as effectively as our symbolic regression model.
    \item \textbf{SVM (SMO)}: This model delivered an accuracy of 80.22\%, Kappa of 0.479, and an F1-score of 0.841. Although SVMs are known for their robustness in classification tasks, our symbolic regression model outperforms it, particularly in terms of F1-score and Kappa.
    \item \textbf{MLP}: The MLP model performed well with an accuracy of 93.13\%, Kappa of 0.546, and an F1-score of 0.945. It surpasses our model in terms of accuracy and F1-score. However, the complexity and interpretability of MLP models are generally lower compared to symbolic regression, which provides a more transparent and interpretable classification framework.
    \item \textbf{Random Forest}: With an accuracy of 97.56\%, a Kappa of 0.622, and an F1-score of 0.981, Random Forest exhibits the highest performance across all metrics. It is particularly strong in sensitivity and specificity, indicating its robustness in handling diverse features and classifying astronomical objects accurately.
\end{itemize}

\subsection{Implications and Interpretability}

One of the key advantages of our symbolic regression approach is its interpretability. The symbolic expressions derived from the regression model provide insight into the relationships between different features and their contributions to the classification task. This transparency is crucial for understanding the underlying patterns in astronomical data, which can be obscured in more complex models like deep learning or ensemble methods.

While our model does not outperform Random Forest in raw accuracy and F1-score, it offers a balance between performance and interpretability. For practical applications where understanding the decision-making process is as important as achieving high accuracy, symbolic regression provides valuable advantages.

\subsection{Future Work}

Future research could focus on integrating symbolic regression with other techniques, such as ensemble methods, to leverage the interpretability of symbolic models with the accuracy of more complex classifiers. Additionally, exploring different feature selection and extraction methods might further enhance the performance of our model.

\section*{Acknowledgments}

The authors used language-enhancement tools for grammar and style corrections, ensuring clarity in the manuscript. All scientific content, methodology, and analysis were conducted independently by the authors.

\bibliographystyle{apsrev4-1}

\bibliography{oja_template}

\end{document}